\documentclass[12pt]{article}
\usepackage[dvips]{graphicx}
\usepackage{pdproc}

  \makeatletter 
  \def\@cite#1{[#1]} 
  \makeatother    
\begin{document}

\renewcommand{\thefootnote}{\alph{footnote}}

\title{
Effect of CP violating phases on neutral Higgs boson\\
phenomenology in the MSSM
}

\author{A.G. AKEROYD$^a$, S. KANEMURA$^b$, Y. OKADA$^{a,c}$, E. SENAHA$^{a,c}$}

\address{a: KEK Theory Group, 1-1 Oho, Tsukuba, Ibaraki 305-0801,
Japan}
\address{b: Department of Physics, Osaka University, Toyonaka, 
Osaka 560-0043, Japan}
\address{c: Department of Particle and Nuclear Physics,
the Graduate University for Advanced Studies, Tsukuba, 
Ibaraki 305-0801, Japan}


\abstract{
In the MSSM with complex SUSY parameters we consider a specific
case where the second heaviest Higgs boson $H_2$ is SM like, 
while $H_1$ and $H_3$ are both strongly mixed states of CP odd and
CP even scalar fields. Such a scenario could be probed at a Linear
Collider via the mechanism $e^+e^-\to H_1H_3$.
}

\normalsize\baselineskip=15pt

\section{Introduction}

The numerous parameters of the MSSM allow for additional
CP violating phases beyond that of the Standard Model CKM phase. 
A phenomenological consequence of complex SUSY parameters 
is ``scalar-pseudoscalar mixing'' \cite{Pilaftsis:1998dd}, 
where the neutral Higgs boson
mass eigenstates are linear combinations of both the scalar (CP even)
and pseudoscalar (CP odd) fields. The $3X3$ neutral Higgs boson mass squared 
matrix $M_{N}^2$ takes the following form: 
\begin{equation}
M_{N}^2 = \left(\begin{array}{cc}
M_{S}^2 & M_{PS}^2 \\
{M_{PS}^2} & M_{P}^2 
\end{array}\right)\qquad \\
\end{equation}
where $M_{S}^2$ ($M_{P}^2$)
denotes the 2X2 (1X1) submatrix for the pure CP even (CP odd) entries, 
while $M^2_{PS}$ mixes the CP odd and CP even
scalar fields. In the MSSM, $M^2_{PS}$ is induced at the 1-loop level
and is explicitly given by \cite{Pilaftsis:1999qt}: 

\begin{equation}
M^2_{PS}=  \Bigm( \frac{m_t^4}{v^2} 
\frac{|\mu||A_t|}{32 \pi^2 M_{SUSY}^2}  \Bigm)  \sin\phi_{CP} 
\times f(M_{SUSY},A_t,\mu,\tan\beta)
\end{equation}

Here $\phi_{CP}=arg(A_t\mu)$, and $f$
is a 
dimensionless function of several SUSY parameters. Setting
$\phi_{CP}=0$ ensures $M_{PS}^2=0$, resulting in Higgs mass eigenstates with
definite CP quantum numbers. In order to have appreciable mixing 
($M_{PS}\approx M_Z$) one requires i) Large $|\mu|/M_{SUSY}$ 
and/or large $|A_{t}|/M_{SUSY}$, and ii) moderate to large 
$\sin\phi_{CP}$.

\begin{figure}[ht]
     \begin{center}
     \begin{tabular}{cc}
     \mbox{\includegraphics*[width=9.3cm]{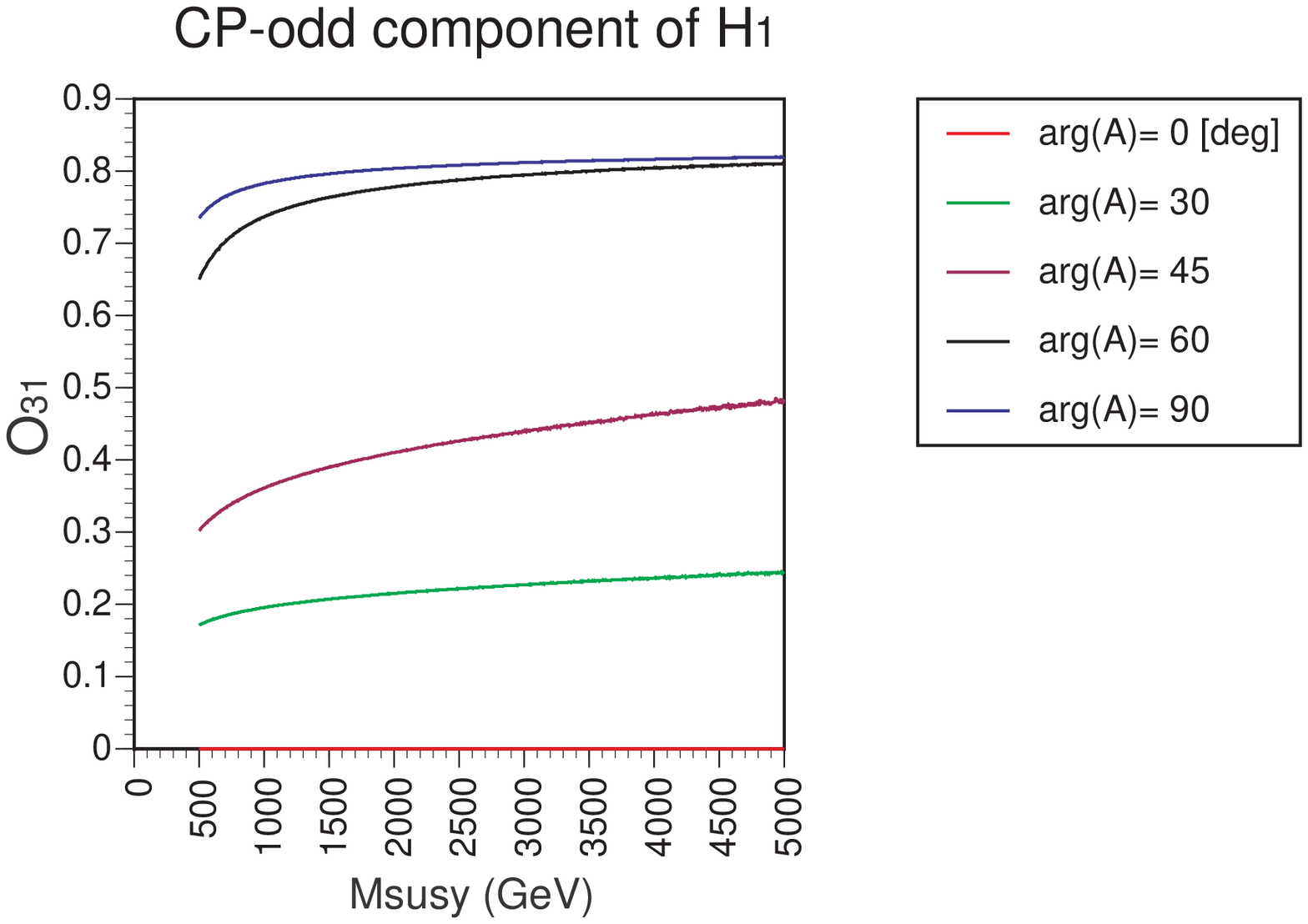}}&
     \mbox{\includegraphics*[width=6.1cm]{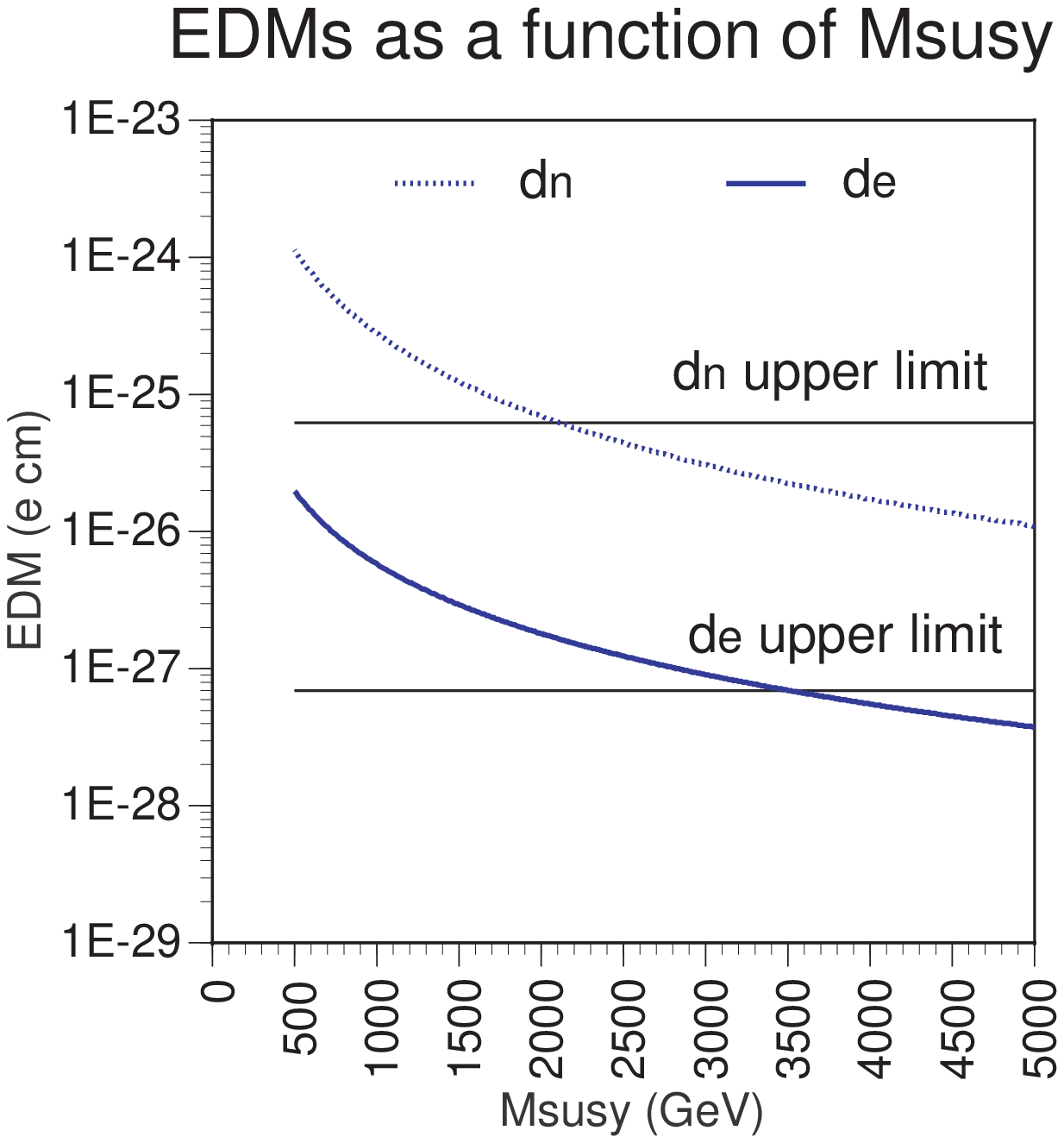}}
     \end{tabular}
     \end{center}
     \caption{Left figure: CP odd composition ($O_{31}$) 
     of the lightest Higgs eigenstate $H_1$ for various
     arg($A_t$) as a function of $M_{SUSY}$; 
     Right figure: Electron and Neutron EDMs for
     arg($A_t=90^\circ$) as a function of $M_{SUSY}$.}
     \label{Fig1}
     \end{figure}

It is known that $\sin\phi_{CP}$ is strongly constrained by fermion 
Electric Dipole Moments (EDMs) if the SUSY spectrum
is relatively light, $M_{SUSY}<1000$ GeV. 
However, the EDMs and $M_{PS}$ exhibit different dependences
on $M_{SUSY}$, provided that the ratios
$|\mu|/M_{SUSY}$ and $|A_{t}|/M_{SUSY}$ are kept fixed
\cite{Ibrahim:2001ht}. This is displayed in 
Fig.~\ref{Fig1} where the CP odd 
component ($O_{31}$) of $H_1$ shows
little sensitivity to $M_{SUSY}$ and can be maximal for large phase,
while the neutron and electron EDMs show a $\sim 1/M^2_{SUSY}$ dependence.

\section{Numerical Results}
We consider a benchmark scenario which gives rise
to a phenomenologically interesting case of strong 
scalar-pseudoscalar mixing which is also compatible with 
the aforementioned EDM constraints.
\begin{eqnarray}
M_{SUSY}=3000\, GeV,\,\, \tan\beta=7\,\, \nonumber \\
|\mu/M_{SUSY}|=10,\,\, |A_t/M_{SUSY}|=1.5,\,\, \phi_{CP}=\pi/2
\end{eqnarray}

    \begin{figure}[ht] 
     \begin{center}
     \vspace*{.2cm}
     \includegraphics*[width=8.5cm]{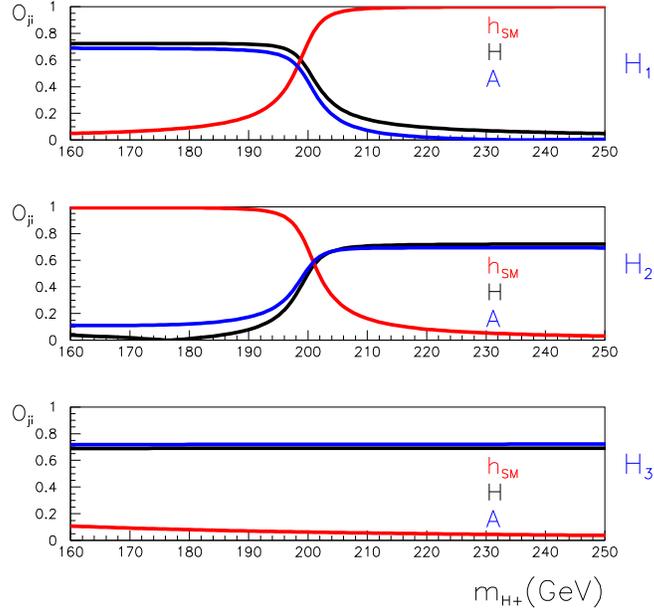}
     \end{center}
     \caption{$O_{ji}$ as a function of $m_{H^\pm}$, for $H_1,
     H_2,H_3$.} 
     \label{Fig2}
     \end{figure}

We work in a rotated
basis in which only one of the $SU(2)XU(1)$ Higgs doublets possesses
a vacuum expectation value, whose corresponding neutral CP
even Higgs boson is given by $h_{SM}$. In this basis $H_i$ are as follows:
\begin{equation}
H_i=O_{1i}H+O_{2i}h_{SM}+O_{3i}A
\end{equation}
In Fig.~\ref{Fig2} we show the composition ($O_{ji}$) of $H_i$
as a function of $m_{H^\pm}$. 
For $m_{H^\pm}<200$ GeV, $H_2$ is SM like and is responsible for 
breaking the $SU(2)XU(1)$ symmetry (see \cite{Carena:2000ks}
for the analogous scenario with smaller $M_{SUSY}$), 
while $H_1$ and $H_3$ are strongly mixed states of $H$ and $A$. For
$m_{H^\pm}> 200$ GeV, one finds the usual decoupling behaviour where
$H_1$ becomes SM like, but the two heavier states $H_2$ and $H_3$ are
strongly mixed states of $H$ and $A$ \cite{Pilaftsis:1999qt}. 
The large mixing for 
$m_{H^\pm}<200$ GeV occurs
because the mass matrix elements for
$H-H,A-A$ and $A-H$ transitions are relatively large and
form a 2X2 submatrix whose entries are roughly equal in magnitude.
The mass matrix elements for $h_{SM}-A$ and $h_{SM}-H$ are
considerably smaller, giving rise to an eigenstate $H_2$ which
is almost purely $h_{SM}$. For $m_{H^\pm}>200$ GeV, the heavy Higgs
$2X2$ submatrix for $A-A,A-H$ and $H-H$ decouples from the
SM like $H_1$. The 
magnitude of the $A-H$ entry is roughly equal to the
difference of the diagonal terms, leading to eigenstates which
are mixed states of CP.

\begin{figure}[ht]
     \begin{center}
     \begin{tabular}{cc}
     \mbox{\includegraphics*[width=8.0cm]{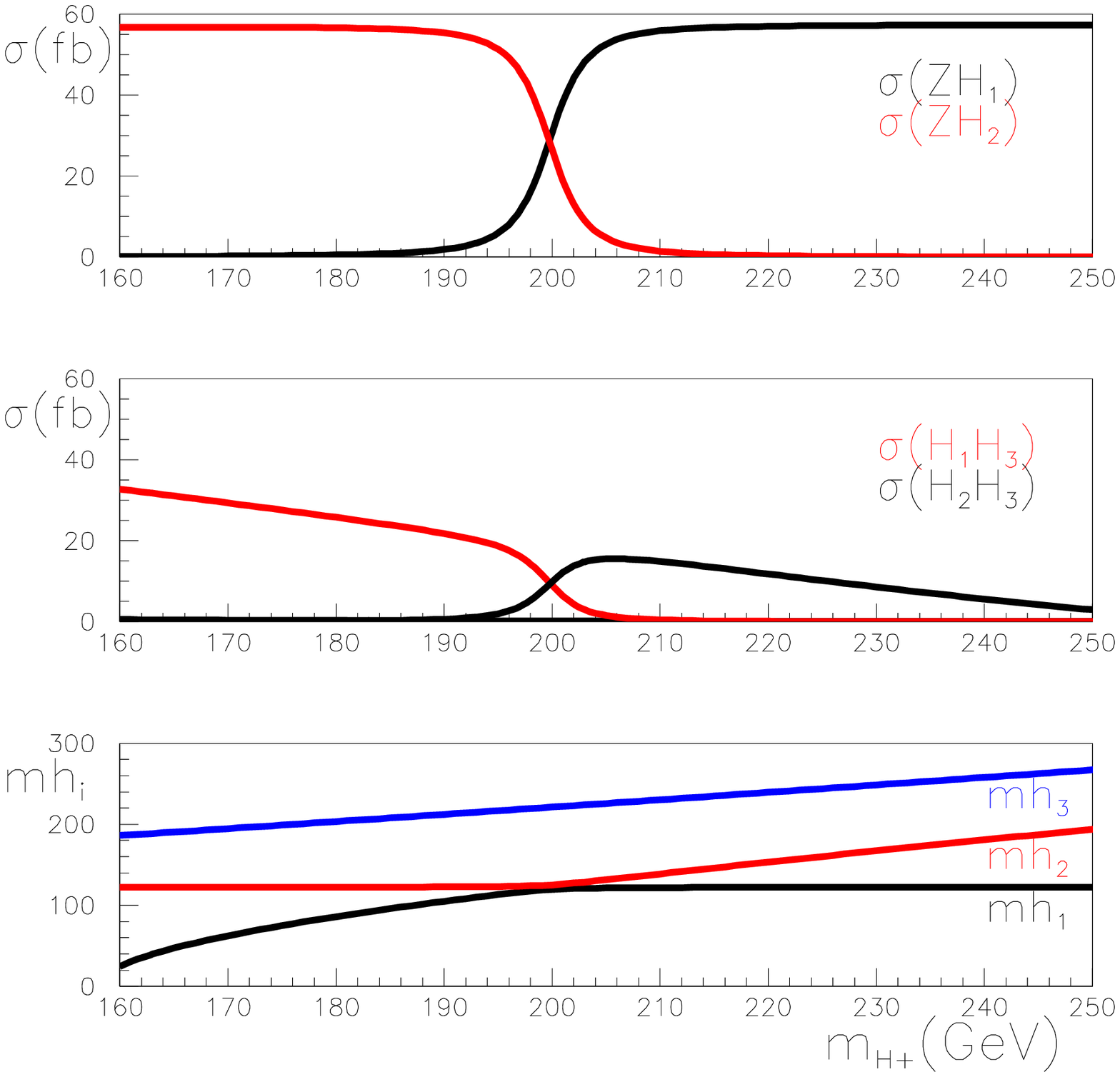}}&
     \mbox{\includegraphics*[width=7.5cm]{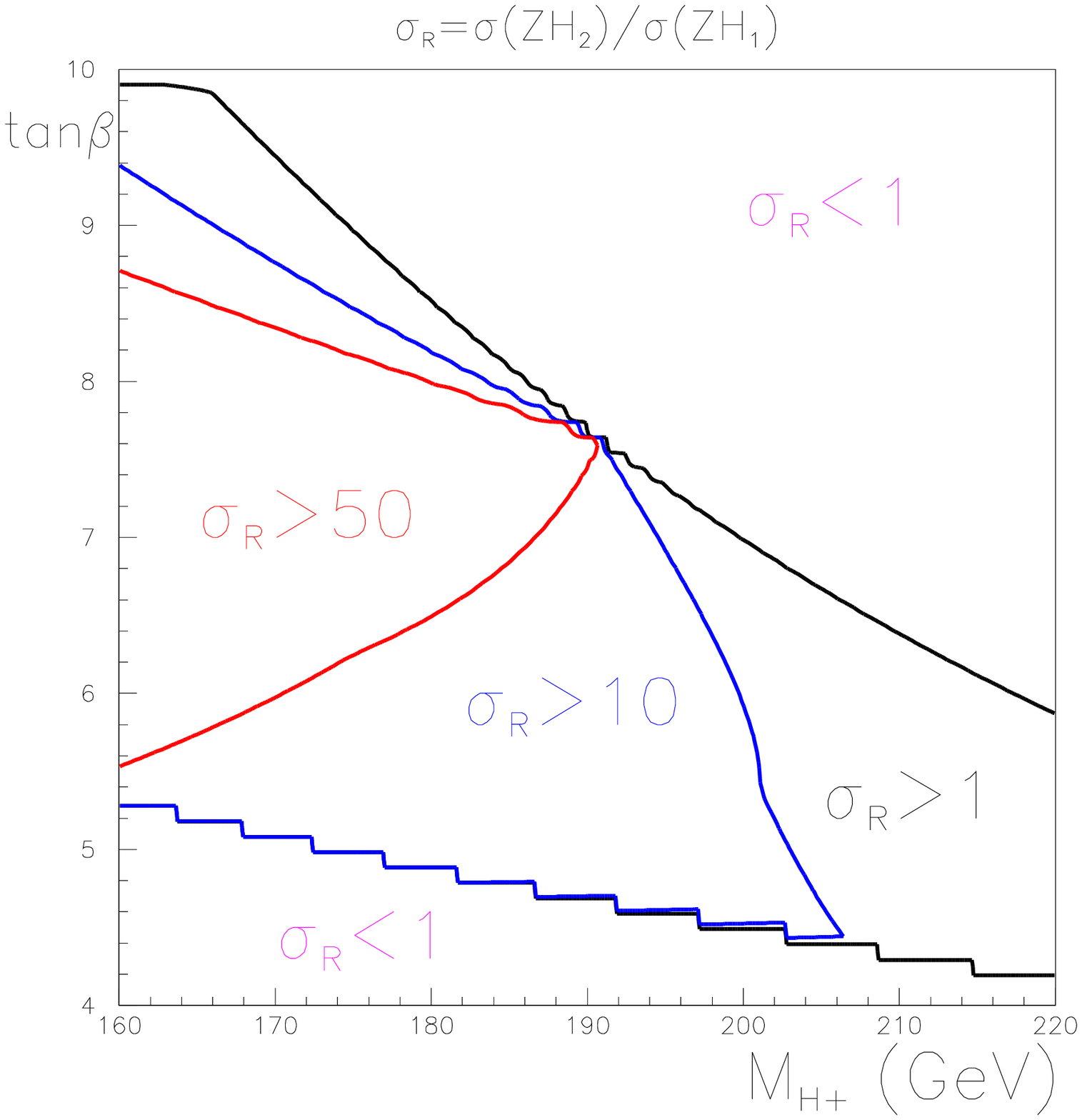}}
     \end{tabular}
     \end{center}
     \caption{Left figure: $\sigma(e^+e^-\to H_iZ)$ (upper), 
     $\sigma(e^+e^-\to H_iH_j$) (middle) and $m_{H_i}$ (lower) as
     a function of $m_{H^\pm}$. Right figure:
     $\sigma_R=\sigma(e^+e^-\to ZH_2)/\sigma(e^+e^-\to ZH_1)$
     in the plane ($m_{H^\pm},\tan\beta$)}
     \label{Fig3}
     \end{figure}

In  Fig.~\ref{Fig3}(left plots) we show $\sigma(e^+e^-\to H_iZ,H_iH_j$) 
as a function of $m_{H^\pm}$, at a Linear Collider of $\sqrt s=500$ GeV.
For $m_{H^\pm}<200$ GeV, $\sigma(e^+e^-\to H_2Z)$
is SM like and thus $H_2$ would be found at the LHC. Detection of
$H_1$ and $H_3$ might be difficult at the LHC, but
$e^+e^-\to H_1H_3$ would offer sizeable rates
at a Linear Collider and is a potentially effective probe of
the scalar-pseudoscalar mixing. Note the large mass splitting
$M_{H_3}-M_{H_1}$ for $M_{H^\pm}< 200$ GeV, a consequence of the  
form of the 2X2 submatrix for the $H-H,A-A,A-H$ entries.
Finally, Fig.~\ref{Fig3} (right plot) shows contours of 
$\sigma_R=\sigma(e^+e^-\to ZH_2)/\sigma(e^+e^-\to ZH_1)$ in the plane
($m_{H^\pm},\tan\beta$). In a sizeable region $\sigma_R > 10$, and thus 
$H_2$ is SM like.

\section{Acknowledgements}
\noindent
This work was supported in part by a Grant-in-Aid of the Ministry
of Education, Culture, Sports, Science and Technology, Government of Japan,
Nos. 13640309 and Nos. 13135225.



\end{document}